\documentclass[conference]{IEEEtran}
\usepackage[utf8]{inputenc}
\usepackage[T1]{fontenc}

\usepackage{graphicx}
\usepackage{subfig}
\usepackage{float}

\usepackage{nth}
\usepackage[noadjust]{cite}
\usepackage[hidelinks]{hyperref}
\usepackage[shortlabels, inline]{enumitem}

\usepackage{array, makecell, multirow, booktabs}
\setcellgapes{2.8pt}

\usepackage{amsmath, amsthm, amssymb, amsfonts, mathtools, optidef}

\theoremstyle{theorem}

\makeatletter
\def\th@definition{
\thm@headfont{\itshape} 
\thm@notefont{} 
{}
}
\makeatother
\theoremstyle{definition}

\usepackage[linesnumbered, algoruled, boxed, lined]{algorithm2e}
\SetKwRepeat{Do}{do}{while} 
\let\oldnl\nl  
\newcommand{\nonl}{\renewcommand{\nl}{\let\nl\oldnl}}  

\usepackage{color, xcolor}

\usepackage{xspace}

\newcommand{\ie}{\textit{i.e.,}\xspace}
\newcommand{\etal}{\textit{et al.}\xspace}

\begin{document}
\bstctlcite{IEEEexample:BSTcontrol}

\title{Unsupervised Data Splitting Scheme for Federated Edge Learning in IoT Networks}

\author{
	\IEEEauthorblockN{Boubakr Nour \qquad Soumaya Cherkaoui}
	\IEEEauthorblockA{Department of Electrical and Computer Science Engineering, Université de Sherbrooke, Canada}
	Email: boubakr.nour@usherbrooke.ca, soumaya.cherkaoui@usherbrooke.ca
}

\maketitle

\begin{abstract}
	Federated Edge Learning (FEEL) is a promising distributed learning technique that aims to train a shared global model while reducing communication costs and promoting users' privacy. However, the training process might significantly occupy a long time due to the nature of the used data for training, which leads to higher energy consumption and therefore impacts the model convergence. To tackle this issue, we propose a data-driven federated edge learning scheme that tends to select suitable participating nodes based on quality data and energy. First, we design an unsupervised data-aware splitting scheme that partitions the node's local data into diverse samples used for training. We incorporate a similarity index to select quality data that enhances the training performance. Then, we propose a heuristic participating nodes selection scheme to minimize the communication and computation energy consumption, as well as the amount of communication rounds. The obtained results show that the proposed scheme substantially outperforms the vanilla FEEL in terms of energy consumption and the number of communication rounds.
\end{abstract}

\begin{IEEEkeywords}
	federated learning, edge computing, data-driven node selection
\end{IEEEkeywords}

\IEEEpeerreviewmaketitle

\section{Introduction}
\label{sec:introduction}
Nowadays, the world is witnessing a substantial development in the field of artificial intelligence (AI), which is sweeping major fields. This development is made because of the tremendous advances in machine learning (ML) techniques. Novel ML applications emerge every day, ranging from autonomous driving and finance to marketing and healthcare, potential applications are boundless~\cite{tang2021survey}. In parallel, wireless mobile technologies, such as the fifth-generation (5G), promise to connect billions of heterogeneous devices to the network edge~\cite{filali2020multi}. This connection opens various opportunities to verticals toward the rise of new applications and business models under the banner of the Internet of Things (IoT)~\cite{laroui2021edge}. Edge nodes seek to collect a massive amount of data and open new avenues for ML applications. The prevalent approach of ML implementations on edge devices is to amass all the relevant data at cloud/edge servers and train powerful ML models using the available data and processing capacities~\cite{nour2021network, liu2020toward}. However, such a centralized fashion is not fitting many use cases' requirements, which might also violate the latency requirements of the underlying application, scatter communication resources, and infringe user/data privacy~\cite{khan2021federated}.
For example, an autonomous car may generate from 5 to 20 TB of data per day. This large volume of data contains sensitive information and needs to be transmitted over the bottleneck core network to be processed at distant cloud servers. The low-latency requirements, information density, and data privacy become challenging to be addressed in such circumstances.

To fulfill the requirements of most IoT applications and promote data privacy, federated edge learning (FEEL)~\cite{malandrino2021federated} has been proposed an advanced machine learning technique to train models in a distributed fashion, using mobile edge computing. In FEEL,
\begin{enumerate*}[(i)]
	\item a centralized edge server shares the model with a set of nodes (\ie participating nodes),
	\item participating nodes train the model locally without sharing their data with the edge server, 
	\item the upload only the model parameters to the edge server,
	\item the latter aggregates the received updates to improve the global model, and
	\item re-sends the model to participating nodes for another training round.
\end{enumerate*}
The process is repeated multiple rounds or until the model reaches a pre-defined threshold.

For each training round, the edge node performs local computation for model training as well as communication with the edge server for model updating. Multiple communication and computation rounds significantly drain the energy of devices that are energy-constrained in nature, and hence restrict their ability to conduct more training rounds. This circumstance negatively impacts the training performance and consequently leads to a slower convergence~\cite{khan2020federated, wang2020optimizing}.
An efficient way to overcome this issue is to reduce the number of training rounds, which reflects to decreasing the computation and communication overhead, and hence, sustain the node's energy. Using fewer data samples for training can help in reaching less communication overhead. Yet, this is a double-edged problem since the model might not converge with fewer data and the accuracy is controversial.
Notwithstanding, work in~\cite{abdellatif2020active} reveals that the model convergence is highly relying on the quality of data used during the training. ML models can be trained with fewer data samples, as long as this data has high quality (\ie variety). Determining the data used in the model training can substantially improve the training performance~\cite{gong2019diversity}.

To this end, in this paper, we design a data-driven federated edge learning scheme for IoT networks. Instead of using the entire dataset for training, edge nodes split their data into a set of sub-datasets by minimizing the data similarity (\ie increase the diversity within each sub-dataset). This approach is motivated by the fact that less yet quality data samples significantly contribute to model learning~\cite{abdellatif2020active}. Thereafter, the edge server selects a subset of nodes to participate in the training based on their similarity score and communication/computation energy.
Overall, the major contributions of this work can be outlined as follows:

\begin{itemize}
	\item we formulate data splitting and node selection problems with the aim to maximize data diversity and minimize energy consumption.
	
	\item we design an unsupervised data-aware splitting scheme that tends to partition local data into a set of sub-datasets with diverse and quality samples.
	
	\item we propose a heuristic node selection scheme that chooses participating nodes with minimum similarity scores while reducing the node's energy consumption.
	
	\item we validate the proposed scheme through simulation using MNIST, a non-independent and identically distributed (non-IID) and unbalanced dataset. The designed scheme helps in model convergence with fewer communication rounds while minimizing energy consumption.
\end{itemize}

The rest of this paper is organized as follows.
Section~\ref{sec:related_work} presents an overview of federated learning and review on related work.
Section~\ref{sec:system_model} introduces the system model and the problem formulation.
Section~\ref{sec:proposed_scheme} details the proposed data-driven federated edge learning scheme.
The performance and effectiveness of the proposed scheme are evaluated in Section~\ref{sec:numerical_results}.
Finally, we conclude the paper in Section~\ref{sec:conclusion}.

\section{Related Work}
\label{sec:related_work}
The integration of artificial intelligence and edge computing drives a mutual benefit that helps to overcome various issues and challenges in today's networks~\cite{wang2020convergence}. For instance, edge computing strengthens from the intelligent decisions made by AI mechanisms, while AI models will be further improved by availing the distributed nature of edge computing. Yet, the edge server requires to collect all data generated by users for training and inference purposes, which might violate the user's privacy~\cite{gu2019privacy}.

Federated learning~\cite{taik2021empowering} aims to overcome the privacy issues in conventional machine learning techniques. In FEEL, multiple edge devices collaboratively train a global model without sharing their data. The process is orchestrated by a centralized edge server that sends the model to multiple nodes. Each node trains the model locally using its own data and sends the model parameters to the server for aggregation. The model is refined and updated for multiple rounds until its convergence~\cite{nour2021federated}.
There have been various efforts in the literature that tend to enhance network resources while ensuring fast model convergence~\cite{yang2021characterizing}. For instance,
authors in~\cite{nishio2019client} propose a client selection scheme for federated learning, taking into account the nature of heterogeneous mobile devices. The edge server sets a deadline to be met by clients, which includes model download, train, update, and upload. The edge server, based on the clients' resources, selects which nodes can participate in the learning phase and hence lower the training time.
Mohammed \etal~\cite{mohammed2020budgeted} tend to minimize the number of IoT candidates contributing to the training process. Inspired by the solution of the secretary problem, the authors design an online heuristic scheme to find the best candidate clients. The objective is to overcome communication and computation constraints while ensuring a predefined accuracy.
Albaseer \etal~\cite{albaseer2021client} propose a client selection approach that leverages devices' heterogeneity. The approach aims to accelerate the convergence rate by using clients' round latency and exploiting the bandwidth reuse for clients that consume more time to update the model. The server performs model averaging and then clusters the clients based on a predefined threshold. Clients with less latency are selected to update the model.
Works in~\cite{albaseer2021fine} and~\cite{albaseer2021threshold} propose a data exclusion approach to minimize energy consumption during federated learning rounds. Toward this, the authors design a customized local training algorithm that selects only the samples that improve the model's quality based on a predefined threshold probability.

Although the aforementioned solutions have contributed to enhancing federated learning architecture, some of them did not consider the data in their studies. Indeed, data is a pillar component when selecting participating clients since it has a direct impact on model accuracy. This work emphasizes data as the pivotal element for model training without ignoring energy consumption.

\section{System Model \& Problem Formulation}
\label{sec:system_model}
\subsection{System Model}
\begin{figure}[!t]
	\centering
	\includegraphics[width=.9\linewidth]{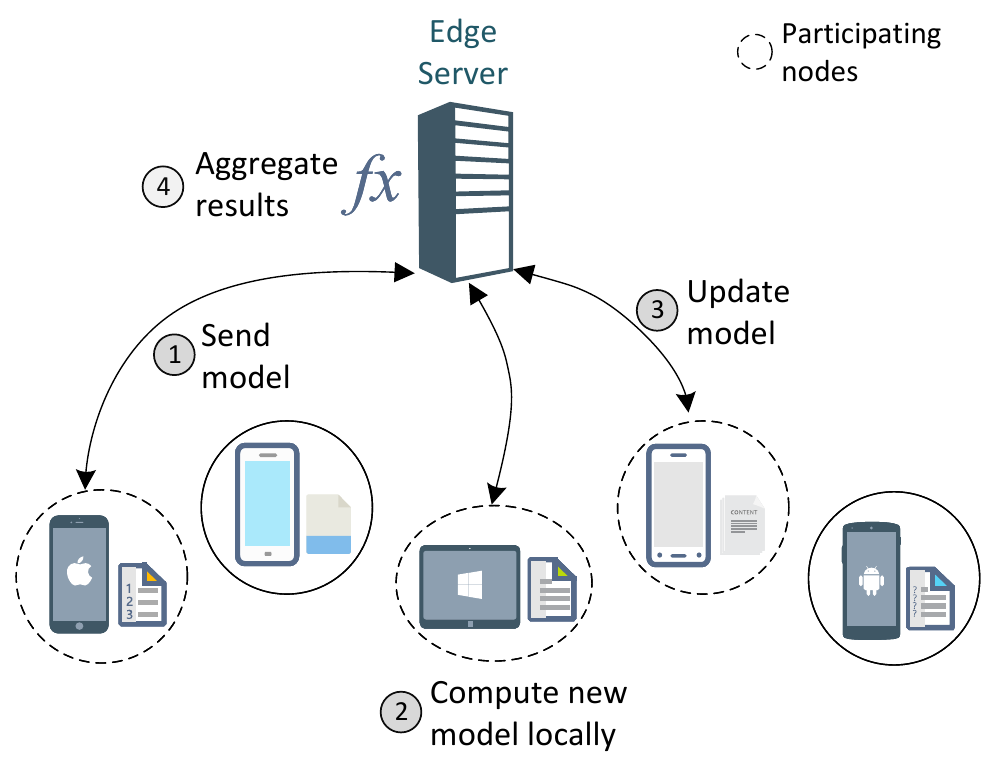}
	\caption{System model for centralized federated edge learning.}
	\label{fig:system_model}
	\vspace{-0.43cm}
\end{figure}

Figure~\ref{fig:system_model} depicts the used system model in this work. Similar to~\cite{albaseer2021threshold}, we consider a network that has a set $\mathcal{K}$ edge devices connected to a centralized edge server. The edge server serves as a coordinator for participating edge nodes to train a global model.

\textit{Learning Model.}
Each edge device $k \in \mathcal{K}$ has its own data $\mathcal{D}_k$ used to train the model locally and sends the update $\omega_k$ back to the edge server. Let $|\mathcal{D}_k|$ denote the number of local data samples for edge device $k$.
The edge server selects a subset of edge devices, denoted $\mathcal{S} \subset \mathcal{K}$ to participate in model training. Those devices receive the initialization parameters $\omega_0$ from the edge server (Step 1), train the model locally using local data $\mathcal{D}_k$ (Step 2), and send the updates to the edge server (Step 3). The edge server, upon receiving the updates, aggregates the received parameters ((Step 4), generates a new global model, and sends the updated model parameters $\omega_r$ to the participating edge devices for another training round. The process is repeated $r_{max}$ rounds until the model converges or reaches a predefined threshold.

Each participating edge device tends to minimize the local loss function. Let $F_{k}^r$, defined in Eq.~(\ref{eq:loss_function}), denote the local loss function for the $k$-th participating device at the $r$-th round:

\begin{equation} \label{eq:loss_function}
	F_{k}^{r}(\omega) \triangleq \frac{1}{|\mathcal{D}_{k}|} \sum_{s \in \mathcal{D}_{k}} f_{s}(\omega).
\end{equation}

\noindent where $f_{s}$ is the error for each local sample in local data. Let $D \triangleq \sum_{k=1}^{|\mathcal{S}|}|\mathcal{D}_{k}|$ denote the total amount of data samples used in edge network, while $\delta_{k} = \frac{|\mathcal{D}_{k}|}{D}$ denote the weight of the local data samples at the $k$-th device.

Edge devices use stochastic gradient descent (SGD)~\cite{wu2020federated} to minimize the loss function, Eq.~(\ref{eq:loss_function}), for $\epsilon$ epochs. The local model parameters $\omega_{k}$ are updated accordingly as shown in Eq.~(\ref{eq:local_model_par}).

\begin{equation} \label{eq:local_model_par}
	\omega_{k}(n) = \omega_{k}(n-1) - \eta \nabla F_{k}^{r}(\omega_{k}(n)).
\end{equation}

\noindent where $n = 1, 2, \dots, \epsilon$ is the number of local updates performed by the $k$-th device and $\eta$ is the learning rate at each round. $\omega_{k}(0)$ is the initial global parameters sent by the edge server and received by the $k$-th device, and $\omega_{k}(\epsilon)$ is the last local parameters update sent by the $k$-th device to the edge server after $\epsilon$ rounds.

After receiving all updates from all participating edge devices, the weighted global loss function at the $r$-th round is calculated as shown in Eq.~(\ref{eq:weighted_global_loss_func}).

\begin{equation} \label{eq:weighted_global_loss_func}
	F_{r}(\omega) \triangleq \sum_{k=1}^{|\mathcal{K}|} \delta_{k} F_{k}^{r}(\omega).
\end{equation}

Accordingly, the global model parameters are calculated as shown in Eq.~(\ref{eq:global_model_par}).

\begin{equation} \label{eq:global_model_par}
	\omega_{r} = \sum_{k=1}^{|\mathcal{K}|} \delta_{k}~\omega_{k},
\end{equation}

Then, both $F_{r}(\omega)$ and $\omega_{r}$ are sent to all participating devices to be employed for the next round ($r + 1$), aiming to minimize $F(\omega)$.

\begin{equation}
	\overset{*}{\omega} \triangleq \arg \min F(\omega).
\end{equation}

\textit{Local Computation Model.}
The participating device randomly partitions local data $\mathcal{D}_k$ into multiple batches of size $b$ and trains the model for $\epsilon$ epochs~\cite{bakhtiari2020federated}. Assuming that the device uses the entire data for training, the local computation time $T_{k}^{\operatorname{cmp}}$ is defined as shown in Eq.~(\ref{eq:local_cmp_time}).

\begin{equation} \label{eq:local_cmp_time}
	T_{k}^{\operatorname{cmp}} = \epsilon \frac{|\mathcal{D}_k| \Phi}{f_{k}^{\operatorname{cmp}}},
\end{equation}

\noindent where $f_{k}^{\operatorname{cmp}}$ denotes the local CPU frequency and $\Phi$ is the number of cycles required to process one data sample.

In order to provide an efficient synchronization of all updates and avoid long waiting time, the edge server sets a deadline $\mathcal{T}$ to send the update back to him. Edge devices have to accomplish the computation and computation within $\mathcal{T}$. Hence, the local computation time is defined as shown in Eq.~(\ref{eq:local_cmp_time_deadline}).

\begin{equation} \label{eq:local_cmp_time_deadline}
	T_{k}^{\operatorname{cmp}} = \mathcal{T} - T_{k}^{\operatorname{up}}.
\end{equation}

\noindent where $T_{k}^{\operatorname{up}}$ is the required time to upload the update to the edge server.

\textit{Energy Consumption Model.}
The energy consumption for the $k$-th device is based on the training process and defined as shown in Eq.~(\ref{eq:local_energy}).

\begin{equation} \label{eq:local_energy}
	E_{k}^{\operatorname{cmp}} = \frac{\alpha_{k}}{2} (f_{k}^{\operatorname{cmp}})^{3} T_{k}^{\operatorname{cmp}},
\end{equation}

\noindent where $\frac{\alpha_{k}}{2}$ is the device's energy capacitance coefficient.
From Eq.~(\ref{eq:local_cmp_time}) and Eq.~(\ref{eq:local_energy}), we find the local energy consumption as:

\begin{equation} \label{eq:local_energy_consumption}
	E_{k}^{\operatorname{cmp}} = \frac{\alpha_{k}}{2} (\varepsilon(f_{k}^{\operatorname{cmp}})^{2} |D_{k}| \Phi).
\end{equation}

Considering a Time Division Multiple Access (TDMA), the model upload latency is defined as in Eq.~(\ref{eq:upload_latency}).

\begin{equation} \label{eq:upload_latency}
	T_{k}^{\operatorname{up}} = \frac{\xi}{\Gamma_{k}}.
\end{equation}

\noindent where $\xi$ denotes the model size and $\Gamma_{k}$ the uplink data rate achieved by the $k$-th device. Therefore, the  transmission energy consumption of the $k$-th device is defined based on the upload latency $T_{k}^{\operatorname{up}}$ and the transmit power $P_{k}^{\operatorname{up}}$, as shown in Eq.~(\ref{eq:transmission_energy_consumption}).

\begin{equation} \label{eq:transmission_energy_consumption}
	E_{k}^{\operatorname{up}} = T_{k}^{\operatorname{up}} P_{k}^{\operatorname{up}}.
\end{equation}

\subsection{Problem Formulation}

\textit{Maximize the data diversity}: the first objective is to accelerate the training phase~\cite{abdellatif2020active}. In doing so, we tend to build sub-datasets with diverse yet quality samples. Therefore, the objective of Eq.~(\ref{eq:obj_sim}) aims at minimizing the similarity between selected samples within the same sub-dataset.

\begin{equation} \label{eq:obj_sim}
	\textbf{P1}: \operatorname{minimize}
	\sum\limits_{\substack{\forall d \in \mathcal{D}\\ \forall u \in d\\ \forall u \neq v \in D}}
	x_u^d x_v^d\operatorname{Sim}(u, v),
\end{equation}

subject to,

\begin{subequations}
	\begin{equation} \label{eq:cons1-1}
		\sum_{u \in d} x_{u}^{d} = 1, \qquad \forall d \in \mathcal{D},
	\end{equation}
	\begin{equation} \label{eq:cons1-2}
		\sum_{d \in \mathcal{D}} x_{u}^{d}, \geq 2 \qquad \forall u \in d,
	\end{equation}
	\begin{equation} \label{eq:cons1-3}
		\sum_{d \in \mathcal{D}}\sum_{u \in d} x_{u}^{d} = |D|,
	\end{equation}
	\begin{equation} \label{eq:cons1-4}
		x_u^d \in \{0, 1\}, \quad \forall u \in d, \forall d \in \mathcal{D}.
	\end{equation}
\end{subequations}

\noindent where $x_i^d$ is a variable indicating the assignment of the $i^{\text{th}}$ sample to the sub-dataset $d$.
Constraints (\ref{eq:cons1-1}) and (\ref{eq:cons1-2}) setup up the outliers-free model where each data sample should be assigned to only one sub-dataset and each sub-dataset should contain at least 2 samples.
Constraint (\ref{eq:cons1-3}) imposes that all data samples should be used to construct the sub-dataset, which implies the fairness in using all data samples in the learning phase.
Finally, the non-negativity constraint is shown in (\ref{eq:cons1-4}).

\textit{Nodes selection}: the second objective (\ref{eq:obj_enrg}) tends to select the appropriate nodes that have qualified data in the training phase while minimizing the energy consumption.

\begin{equation} \label{eq:obj_enrg}
	\textbf{P2}: \operatorname{minimize}
	\sum\limits_{\substack{\forall k \in \mathcal{K}}}
	\gamma_k (E_{k}^{\operatorname{cmp}} + E_{k}^{\operatorname{up}}),
\end{equation}

subject to,

\begin{subequations}
	\begin{equation} \label{eq:cons2-1}
		\gamma_{k} (T_{k}^{\operatorname{cmp}} + T_{k}^{\operatorname{up}}) \leq \mathcal{T}, \quad \forall k \in \mathcal{K},
	\end{equation}
	\begin{equation} \label{eq:cons2-2}
		\gamma_{k} (E_{k}^{\operatorname{cmp}} + E_{k}^{\operatorname{up}}) \leq E_{k}, \quad \forall k \in \mathcal{K},
	\end{equation}
	\begin{equation} \label{eq:cons2-3}
		\gamma_{k} \in \{0, 1\}, \quad \forall k \in \mathcal{K}.
	\end{equation}
\end{subequations}

\noindent where $\gamma_{k}$ indicates the node selection variable. $\gamma_{k} = 1$ if the node $k$ has been selected in the training phase, otherwise $\gamma_{k} = 0$.
Constraint (\ref{eq:cons2-1}) imposes that the computation and computation should be accomplished before the fixed deadline by the edge server.
Similarly, constraint (\ref{eq:cons2-2}) enforces that computation and transmission energy should not exceed the node available energy. This constraint helps in ensuring that the selected node is able to provide training and successfully send the update to the edge service without its power running off.
Finally, constraint (\ref{eq:cons2-3}) expresses the non-negativity.

\begin{algorithm}[!t]
	\SetKwInput{KwData}{Input}
	\KwData{\\
		~~$D$: local dataset\;
		~~$k$: number of sub-datasets\;
	}
	\SetKwInput{KwResult}{Output}
	\KwResult{\\
		~~$\{\mathcal{D}_i, \dots, \mathcal{D}_k\}$: sub-datasets\;
	}
	
	\BlankLine
	$R$ $\leftarrow D$.random($k$)\;
	$i$ $\leftarrow$ 1\;
	\For{$r \in R$}{
		$\mathcal{D}_i$.append($r$)\;
		$r$.marked $\leftarrow$ True\;
		$i \leftarrow i + 1$\;
	}
	$Q \leftarrow D - R$\;
	
	\BlankLine
	\While{($|Q|$ > 0)}{
		$v \leftarrow Q$.random()\;
		
		\If{($v$.marked $\leftarrow$ False)}{
			\For{$i \in [1,~k]$}{
				\While{($e \leq |\mathcal{D}|$)}{
					$\mathcal{S}(\mathcal{D}_i) \leftarrow \operatorname{Sim}(\mathcal{D}(e), v$)\;
					$e \leftarrow e + 1$\;
				}
				$j \leftarrow \operatorname{min}(\mathcal{S}(\{\mathcal{D}_i, \dots, \mathcal{D}_k\})$)\;
				$\mathcal{D}_j$.append($v$)\;
				$v$.marked $\leftarrow$ True\;
			}
		}
	}
	\Return $\mathcal{D}$
	
	\caption{Unsupervised Data-Aware Splitting Scheme.}
	\label{alg:data_splitting}
\end{algorithm}

\section{Data-Driven Federated Edge Learning}
\label{sec:proposed_scheme}
In order to solve the aforementioned problems, we design unsupervised data-aware splitting and heuristic node selection schemes. The unsupervised data-aware splitting scheme runs at the edge device and enables the node to split its local dataset into a set of sub-dataset that include quality samples by maximizing the data diversity. The node selection scheme, on the other hand, is executed at the edge server to select participating nodes, that own quality data, in the training phase with minimum energy consumption. The joint goal is to significantly enhance the global model training and improve the overall network performance.

\subsection{Data-Aware Splitting Scheme}
Algorithm~\ref{alg:data_splitting} shows the process of a data-aware splitting scheme that tends to solve \textbf{P1}. The algorithm allows the edge device to efficiently split its large-size dataset into a set of sub-datasets. The splitting process needs to maximize the data diversity so that the model can converge quickly with fewer data samples.

In doing so, each node independently selects a number ($k$) of desirable sub-datasets to be created, randomly picks $k$ samples from the original data, and then assigns each one to each sub-dataset (lines 1-7).
Afterward, and for each sample $v$ in the remaining data, if $v$ has not been assigned to any sub-dataset, the scheme calculates the similarity index of the said sample versus each sub-dataset (lines 13-16). At the end of the process, the sample is assigned to the sub-dataset with the minimum similarity score. 
The algorithm returns a list of $k$ sub-datasets that has minimum similarity scores. 

In this work, we apply Jaccard Index to calculate the similarity between data samples which is measured as shown in Eq.~\ref{eq:jaccard_index}, where $d_i$ and $d_j$ are two data samples. An index of 0 means that samples are completely dissimilar, an index of 1 means that they are identical, and an index between 0 and 1 represents a degree of similarity.

\begin{equation} \label{eq:jaccard_index}
	J(d_i, d_j) = \frac{d_i \cap d_j}{d_i \cup d_j}.
\end{equation}

\subsection{Participating Node Selection Scheme}
To solve \textbf{P2}, we design a heuristic nodes selection scheme. After the edge server sends participation requests, nodes willing to participate in the training phase send information, such as the number of created sub-datasets with their size and similarity scores, available energy, and communication and computation energy, etc., back to the edge server. It is important to highlight that this information does not carry any sensible input that violates the data/user privacy. Algorithm~\ref{alg:node_selection} shows the overall process executed by the edge server. 

\begin{algorithm}[!t]
	\SetKwInput{KwData}{Input}
	\KwData{\\
		~~$N$: list of edge nodes\;
		~~$k$: number of required participating nodes\;
	}
	\SetKwInput{KwResult}{Output}
	\KwResult{\\
		~~$\mathcal{S}$: list of selected participatig nodes\;
	}
	
	\BlankLine
	Send participation initialization\;
	Nodes execute Algorithm 1\;
	Nodes reply with available information\;
	Assign coefficients to nodes\;
	Order nodes based on similarities and energy\;
	Pick the first $k$ nodes\;
	\BlankLine
	
	\For{$r \leq r_{max}$}{
		\For{$c \in \mathcal{K}$}{
			Send the global model\;
			Train model locally\;
			Send updated model with available energy\;
		}
		Aggregate the model\;
		Update nodes information;\
		Order nodes based on similarities and energy\;
		Pick the first $k$ nodes\;
		$r \leftarrow r + 1$\;
	}
	
	\caption{Participating nodes selection scheme.}
	\label{alg:node_selection}
\end{algorithm}

For each node, the edge server assigns a coefficient based on the number of available sub-dataset, similarity index, and energy. Smaller coefficients are assigned to nodes with a minimum similarity index and larger energy. The edge server sorts the information and picks the first $n$ nodes to participate in the learning. The node's coefficient helps the edge server to pick a minimum number of nodes with diverse data samples instead of picking multiple nodes. The smaller number of selected nodes is, the least communication overhead performed.
Since the energy of nodes can be changed over time, the node selection process is repeated for each round. New nodes can be selected for the next training round while already selected nodes could be dropped out.

\section{Numerical Results}
\label{sec:numerical_results}
In the experiment part, we consider a realistic federated edge learning environment, as depicted in Figure~\ref{fig:system_model}, that consists of 100 nodes. We used the MNIST database~\cite{lecun1998gradient}, an imbalanced and non-IID data distribution with different CCN models. Edge devices use mini-batch stochastic gradient descent as a local solver~\cite{wu2020federated}. The global model evaluation is performed every round with a learning rate of $0.001$. The original dataset is randomly partitioned into $N$ pieces where each node is assigned one part. The database is split into 80\% for training and 20\% for testing.
We compared the proposed scheme against vanilla federated edge learning (\ie standard FEEL scheme) and random nodes selection (\ie selecting random participating nodes). Results are collected and shown in an average of 10 trials.

\begin{figure}[!t]
	\centering
	\includegraphics[width=.95\linewidth]{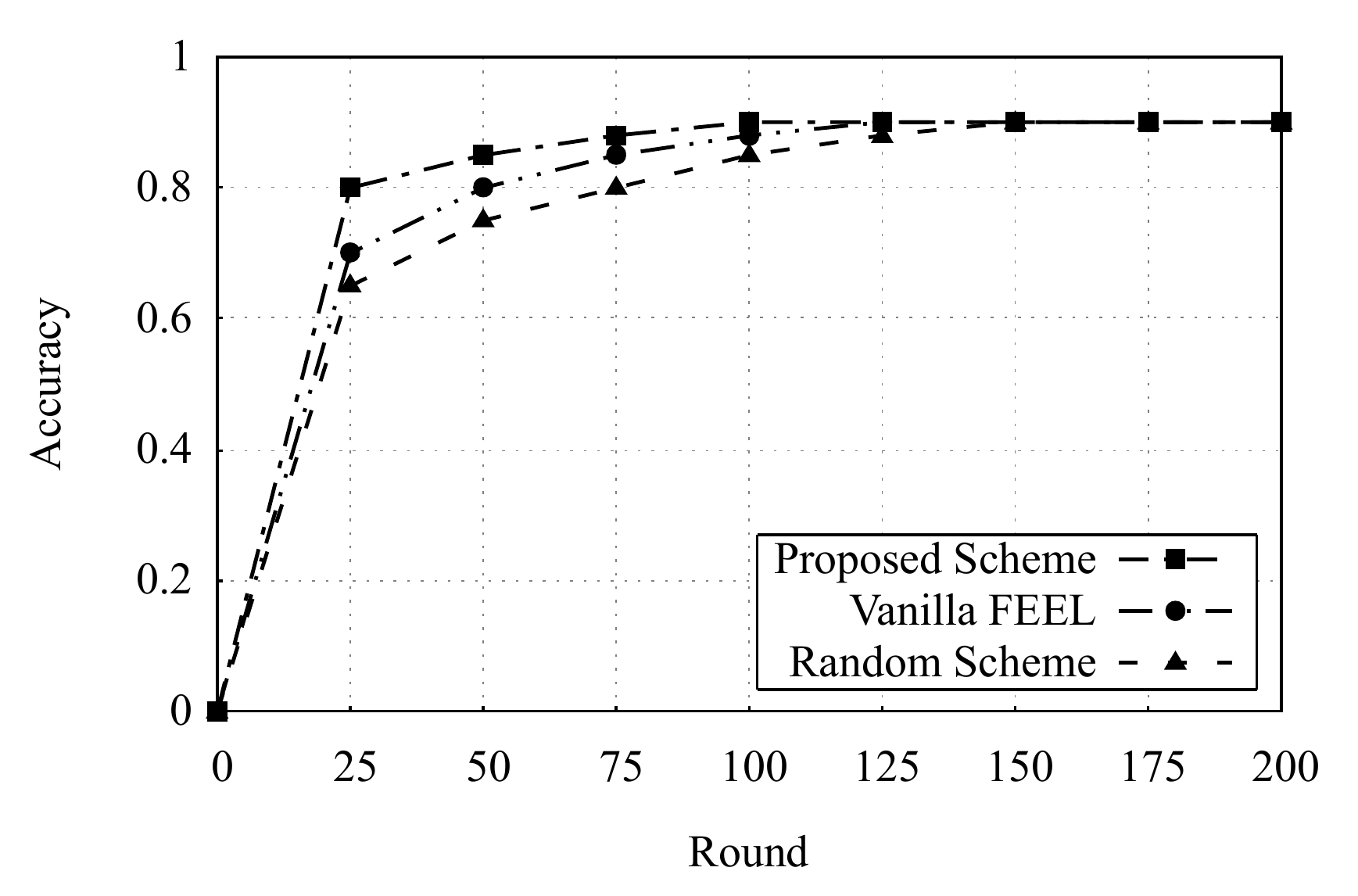}
	\caption{Average accuracy.}
	\label{fig:accuracy}
	\vspace{-0.4cm}
\end{figure}

\begin{figure}[!t]
	\centering
	\includegraphics[width=.95\linewidth]{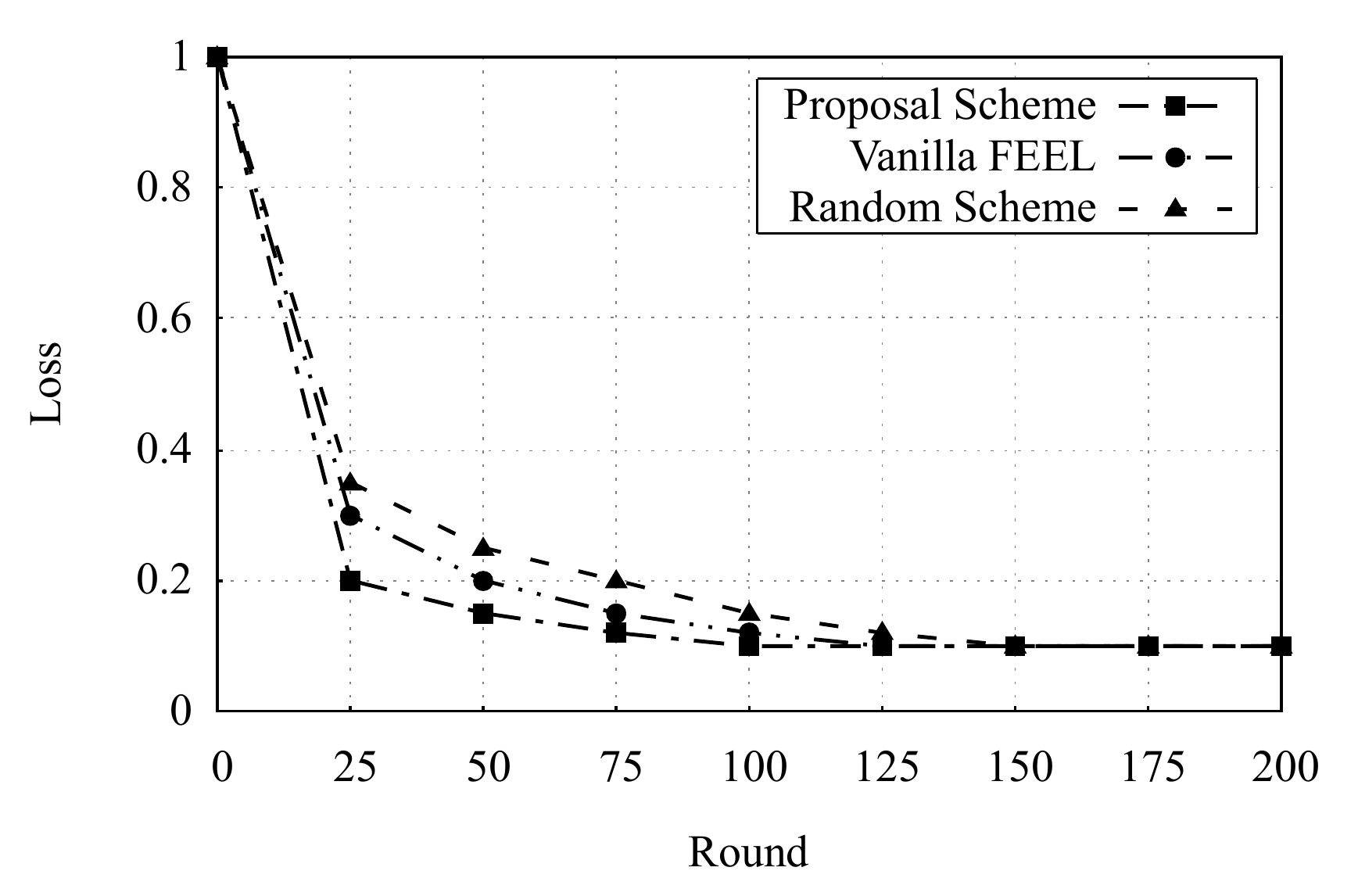}
	\caption{Average loss.}
	\label{fig:loss}
	\vspace{-0.4cm}
\end{figure}

\begin{figure*}[!t]
	\centering
	\begin{minipage}[b]{0.45\linewidth}
		\centering
		\includegraphics[width=\linewidth]{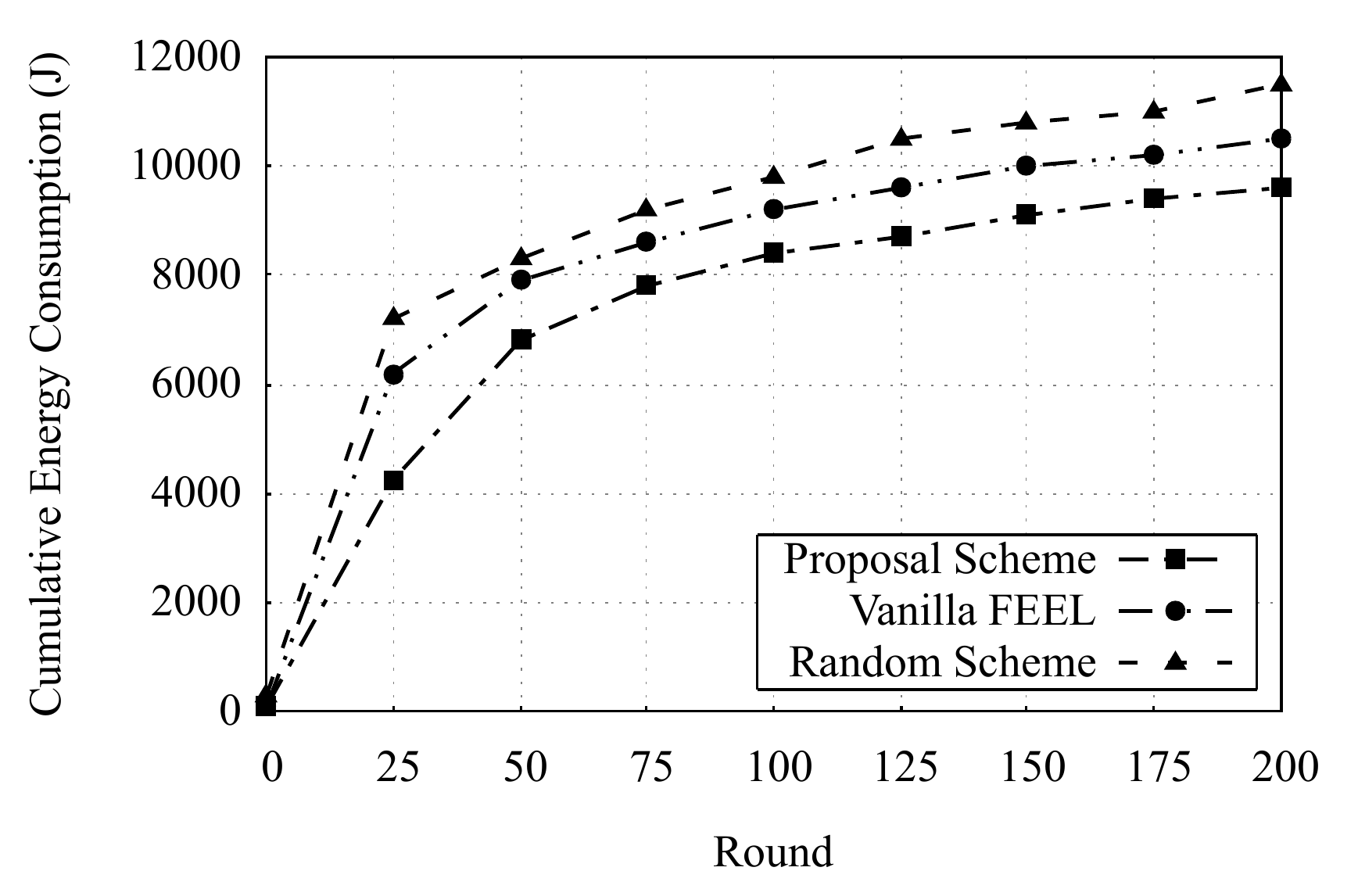}
		\caption{Average cumulative energy consumption.}
		\label{fig:cumulative_energy}
	\end{minipage}
	\hspace{0.01cm}
	\begin{minipage}[b]{0.45\linewidth}
		\centering
		\includegraphics[width=\linewidth]{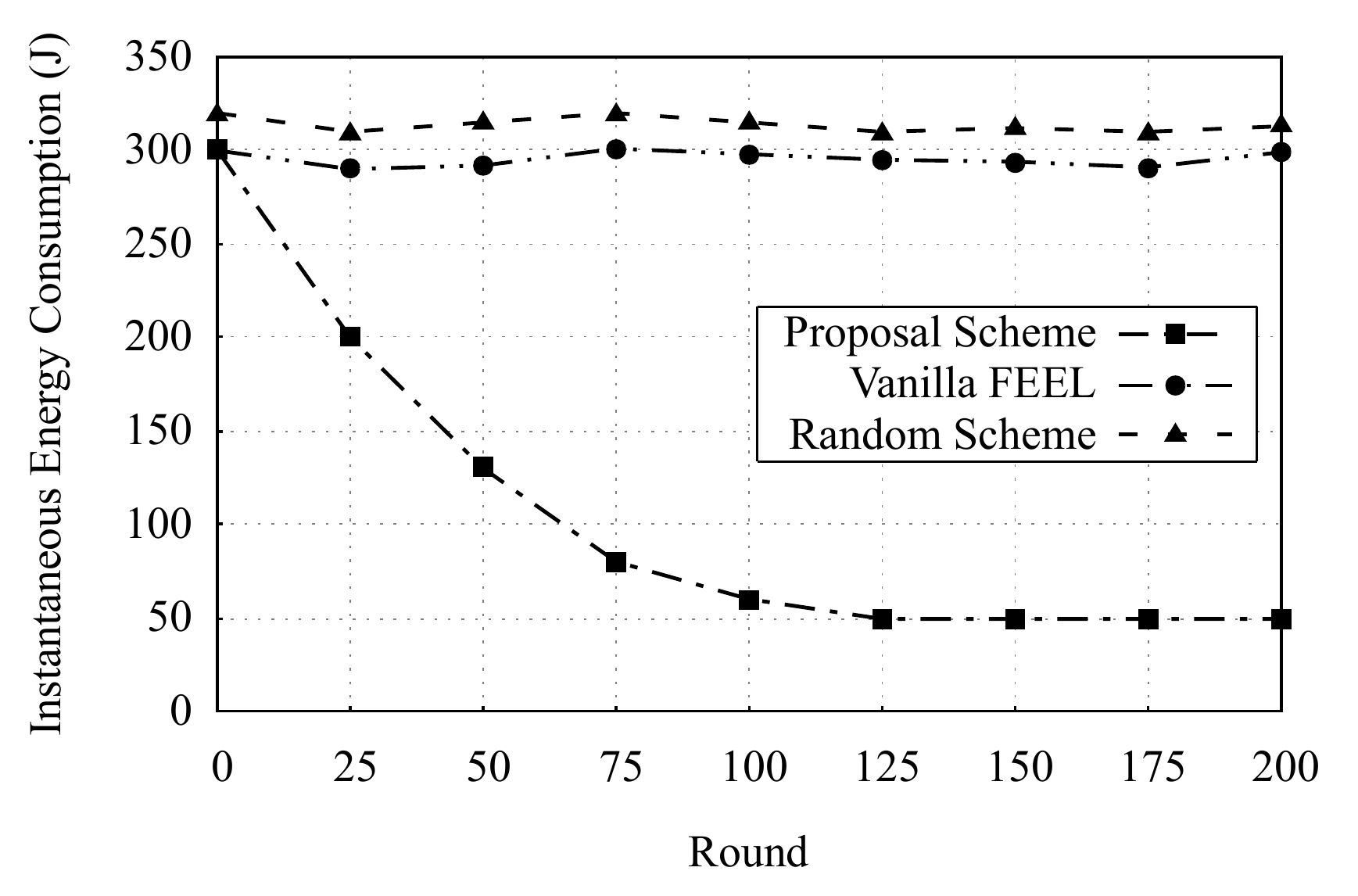}
		\caption{Average instantaneous energy consumption.}
		\label{fig:instantaneous_energy}
	\end{minipage}
	\vspace{-0.4cm}
\end{figure*}

Figures~\ref{fig:accuracy} and~\ref{fig:loss} show the evaluation of the accuracy and the loss function for 200 training rounds, respectively. The loss function used is Mean Squared Error. We can see that the proposed scheme is not affected by reducing the number of participating nodes or the number of samples used in the learning. Indeed, the obtained accuracy and loss are better than the vanilla FEEL and the random scheme. With a fewer number of participating nodes and a small size of data, the proposed scheme was able to converge quickly.

Figures~\ref{fig:cumulative_energy} and~\ref{fig:instantaneous_energy} present the cumulative and instantaneous energy consumption results, respectively. We can observe that the proposed scheme shows a significant reduction in energy consumption compared with vanilla FEEL and random node selection scheme. The reduction gain is proportional to the number of training rounds. This is due to the fact that the proposed scheme selects a minimum number of nodes to participate in the training with diverse and quality data samples. The diversity helps in fast model convergence with fewer data and training rounds. Both vanilla FEEL and random schemes need to incorporate the entire dataset that requires more time for training, computation, and communication and by consequence more energy consumption.

\section{Conclusion}
\label{sec:conclusion}
In this paper, we designed a data-driven federated edge learning scheme for IoT networks. Our design goal is to enhance the model convergence with the minimum energy consumption. In doing so, we proposed an unsupervised data-splitting scheme that partitions the node's local data into diverse and quality sub-datasets to be used in the training. We applied the Jaccard index to ensure the diversity of data samples. We also proposed a heuristic node selection scheme to select participating nodes with minimum energy consumption. The experimental evaluation demonstrates the effectiveness and efficiency of the proposed scheme in terms of model convergence and energy consumption. As a future step, we will focus on designing a distributed data-aware exclusion and node scheduling architecture for federated learning.

\section*{Acknowledgments}
The authors would like to thank the Natural Sciences and Engineering Research Council of Canada (NSERC) for the financial support of this research.

\bibliographystyle{IEEEtran}

\end{document}